# A multi-objective synthesis methodology for majority/minority logic networks

**Moein Sarvaghad-Moghaddam[1], Ali A. Orouji*[1] and Monireh Houshmand[2]**

[1] Department of Electrical and Computer Engineering, Semnan University, Semnan, Iran

[2] Department of Electrical Engineering, Imam Reza International University, Mashhad, Iran

*Email: aliaorouji@semnan.ac.ir

**Abstract**
New technologies such as Quantum-dot Cellular Automata (QCA), Single Electron Tunneling (SET), Tunneling Phase Logic (TPL) and all-spin logic (ASL) devices have been widely advocated in nanotechnology as a response to the physical limits associated with complementary metal oxide semiconductor (CMOS) technology in atomic scales. Some of their peculiar features are their smaller size, higher speed, higher switching frequency, lower power consumption, and higher scale integration. In these technologies, the majority (or minority) and inverter gates are employed for the production of the functions as this set of gates makes a universal set of Boolean primitives in these technologies. An important step in the generation of Boolean functions using the majority gate is reducing the number of involved gates. In this paper, a multi-objective synthesis methodology (with the objective priority of gate counts, gate levels and the number of inverter gates) is presented for finding the minimal number of possible majority gates in the synthesis of Boolean functions using the proposed Majority Specification Matrix (MSM) concept. Moreover, based on MSM, a synthesis flow is proposed for the synthesis of multi-output Boolean functions. To reveal the efficiency of the proposed method, it is compared with a meta-heuristic method, multi-objective Genetic Programing (GP). Besides, it is applied to synthesize MCNC benchmark circuits. The results are indicative of the outperformance of the proposed method in comparison to multi-objective GP method. Also, for the MCNC benchmark circuits, there is an average reduction of 10.5% in the number of levels as well as 16.8% and 33.5% in the number of majority and inverter gates, as compared to the best available method respectively.

## 1 Introduction

The requirements for increasing speed and decreasing power have led to scaling of feature sizes in Complementary Metal–Oxide–Semiconductor (CMOS) technology. More scaling of feature sizes is not possible due to physical limits such as quantum effects and non-deterministic behavior of small currents [1]. Hence, in response to the mentioned limitations, a number of other methods such as Quantum-dot Cellular Automata (QCA) [2], Single-Electron Tunneling (SET) [3,4], Tunneling Phase Logic (TPL) [5] and spin-based devices [6-9] can be used as possible alternatives to CMOS.

QCA was first proposed by Lent (1993) [10,11]. QCA is a promising transistor-less technology and beyond-CMOS technology and will play a crucial role in the future of supercomputing [12,13]. The fundamental unit of QCA is a QCA cell which is composed of four dots located at the corners of a square. This technology acts on the basis of Coulombic interactions of electrons trapped in quantum dots. In QCA, the three-input majority and inverter gates are the fundamental primitives.

All-spin logic (ASL) [14] nanotechnology also implements majority logic gates. The fundamental logical device in TPL is a minority gate [5], which is the complement of majority logic. The minority logic synthesis problem is analogous to the majority logic synthesis problem. SET implements both majority and minority logic [4,3].

In CMOS technology, "NAND/NOR/inverter" gates are used to implement circuits; thus, methods created for synthesis of functions such as Karnaugh maps (K-maps), which produce simplified expressions in the two standard forms named as Sum Of Product (SOP) and Product Of Sum (POS), are not efficient enough for synthesis of functions to present the simplest possible form for the QCA technology.

Some of researchers [15-18] have proposed effective solutions to the synthesis of QCA-based logic structures. However, these methods were only suitable for small networks as they were used to manually solve the problems or synthesize three-input functions. Synthesis methods proposed in [19,16] are based on a geometrical interpretation of only three-variable Boolean functions to reduce the majority expressions created by sum of products. These methods led to the creation of thirteen standard functions, which are used for synthesis of the three-input functions. Huo et al. in their study [20] have introduced a table consisting of twenty standard functions and their corresponding majority expressions. First, a given Boolean function is simplified to a function presented in the mentioned table, and then as a result, a majority expression equivalent to this table is chosen. Some methods [21-24] have applied meta-heuristic algorithms such as Genetic Algorithm (GA) and Genetic Programming (GP) for simplification of logic functions. Bonyadi et al. [21] used GA for optimization of a given single-output Boolean function by majority and inverter gates while Houshmand et al. in [22,23] applied GP algorithm for optimization of multi-outputs functions. In [24], the work proposed in [21] has been extended, and a multi-objective optimization consisting of delay as well as the number of gates have been considered. In [25,26], by using the standard functions, Boolean functions decomposed to four-feasible networks were converted to their corresponding majority expressions. However, the standard functions obtained in [25,26] cannot be considered as a complete set. In [27], full set of standard functions, which is not optimal, was identified according to graph theory.

In this paper, a multi-objective synthesis methodology (with the objective priority of gate counts, gate levels, the number of NOT gates) is proposed, which can be used for the synthesis of three, four, or higher input functions. The concept of Majority Specification Matrix (MSM) is introduced and employed. Furthermore, the synthesis flow is considered for synthesizing multi-output functions. Since the synthesized majority networks can be trivially converted into minority networks using De Morgan's theorem, we only focus on majority logic synthesis in this study. To compare the suggested method with the other ones, benchmarks in [24] and MCNC benchmark circuits are used. This approach results in fewer majority gates and fewer logic levels as compared to existing methods [28,29]. The resulting majority/minority network can then be used in ASL-,QCA-, TPL-, or SET-based nanotechnologies.

The rest of the paper is organized as follows: In Section 2, some related background materials are presented. Section 3 introduces the proposed method in detail. In Section 4, a synthesis flow for multi-output functions is introduced. Section 5 presents the results, and finally Section 6 concludes the paper.

## 2 Background material

In this section, basic concepts in QCA technology such as Quantum-dot cellular automata and QCA devices are explained.

### 2.1 Quantum-dot cellular automata

A standard QCA cell (Figure 1.a)) is composed of four dots located at the corners of a square. Two free electrons can tunnel to any quantum-dot within the cell [30]. Because of Coulombic interactions, the electrons occupy diagonally opposite positions. Depending on the position of the cell, polarization of a QCA cell can be determined with two stable cell-polarization states as shown in Figure 1.b). These configurations are denoted as cell polarization $P = +1$ (binary '1' state) and $P = -1$ (binary '0' state).

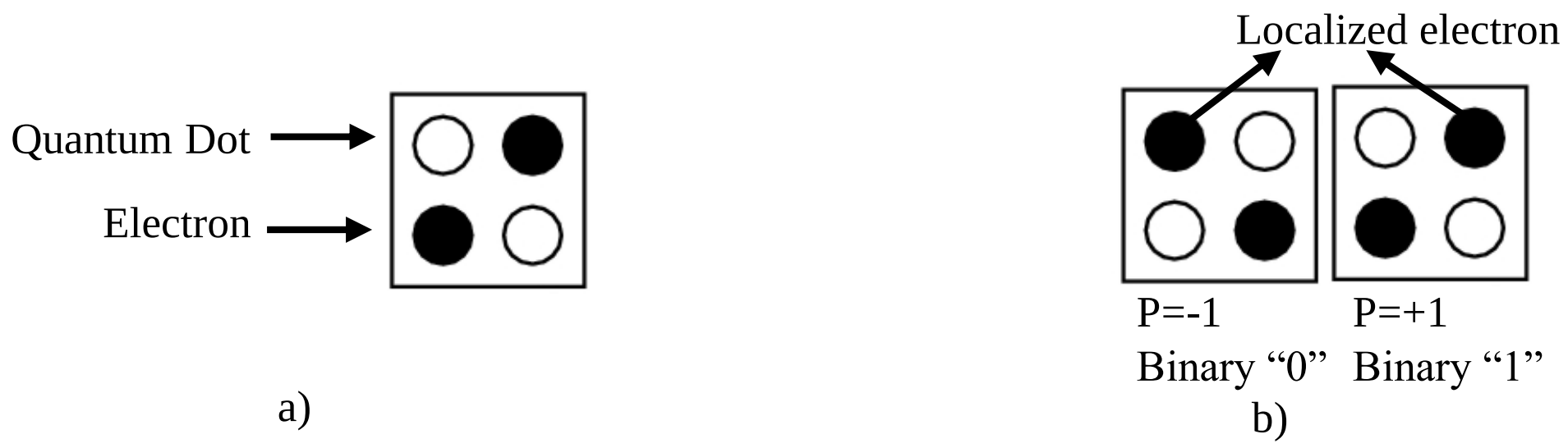

Figure 1: a) Structure of a QCA cell with four quantum dots. b) QCA cell with two different polarizations.

### 2.2 QCA devices

In this section, the basic devices used in QCA such as QCA wires, QCA inverters and QCA majority voters will be introduced. In a QCA wire, a binary signal propagates from input to output because of the Coulombic interactions between cells. In a QCA inverter, cells oriented at 45° to each other take on opposing polarization. A QCA majority gate can perform a three-input majority gate. Equation (1) presents the logic function of a three-input majority gate where *A*, *B*, and *C* are the three inputs.

$$M(A,B,C) = AB + BC + CA. \tag{1}$$

By forcing one of the three inputs of the majority gate to a constant logic "0" or a "1" the majority gate can be used to perform AND/OR operations as shown in the following equations:

$$M(A,B,0) = AB, \qquad M(A,B,1) = A + B. \tag{2}$$

Figure 2 demonstrates a QCA wire (a), inverter gate (b), and majority gate (c), respectively.

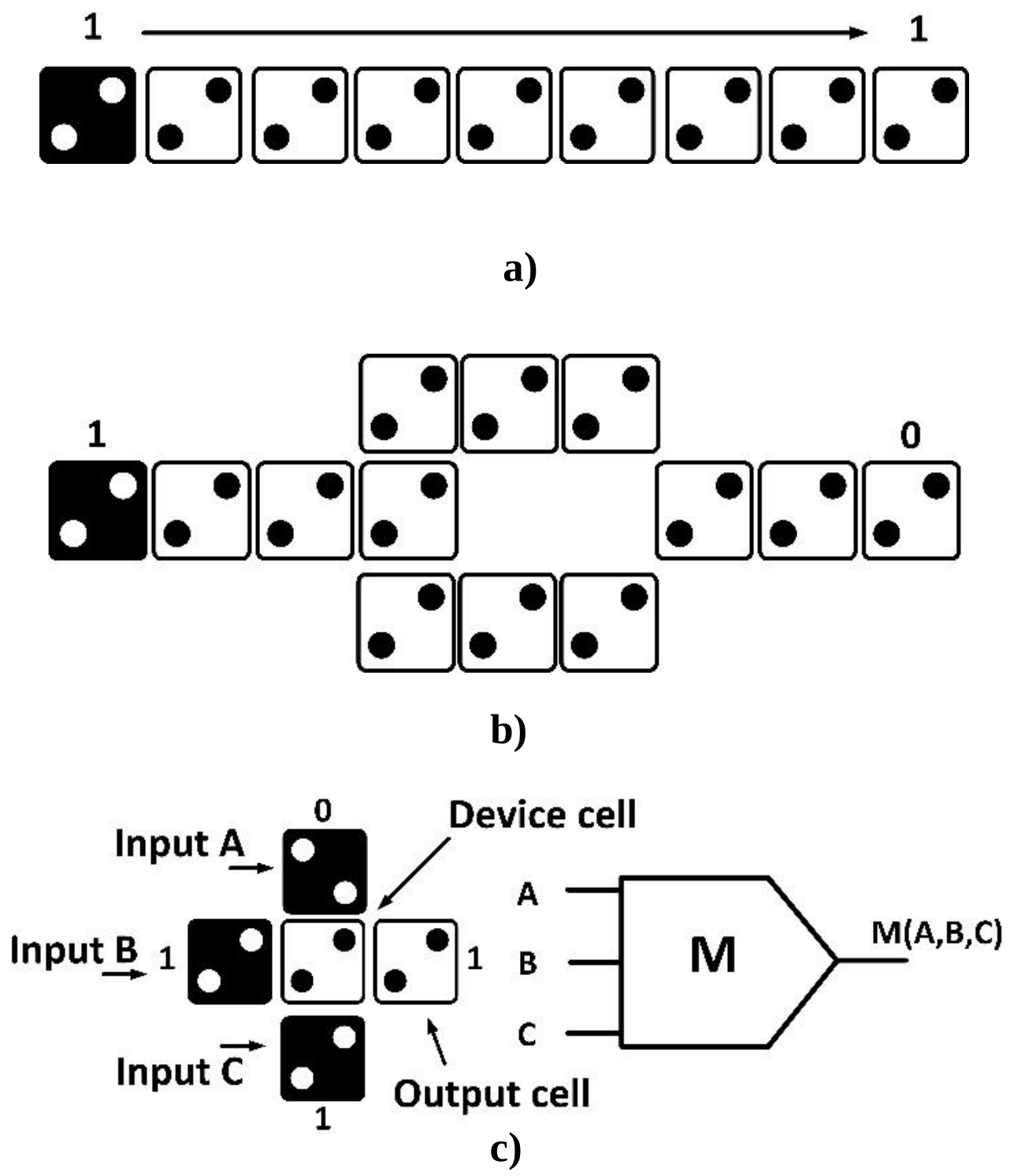

Figure 2: Representation of a) QCA wire b) inverter gate c) QCA majority gate.

## 2.3 Spin devices

All-spin logic (ASL) is a low power switch with switching mechanism based on spin-torque. In these circuits, the input and output are in the electrical domain, while the processing within the circuit happens in the spin domain. Figure 3.a) shows the layout of an ASL device which has four terminals (a) terminal1: VDD, (b) terminal2: VSS, (c) terminal3: input and (d) terminal4: output. The device is composed of one magnet (or nanomagnet), which is the information storage unit, one high polarization ($High_P$) magnet channel interface for the input, one low polarization ($Low_P$) magnet channel interface for the output, an isolation between receiving (input) and transmitting (output) sides and spin-channels both at receiving and transmitting sides as shown in Figure 3.a). The two stable states of the magnet (left and right spin) are determined by the magnet anisotropy (uniaxial anisotropy, Ku) [31]. Also, in Figures 3.b) and c) the ASL inverter and majority gates are shown [6].

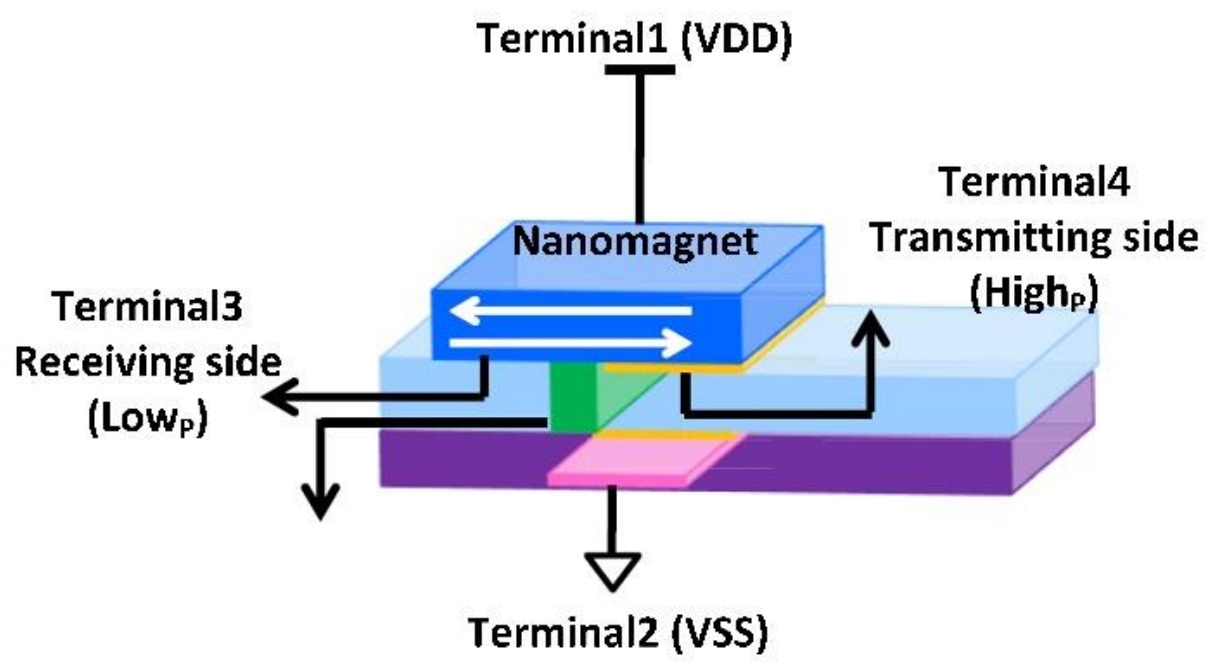

**a)**

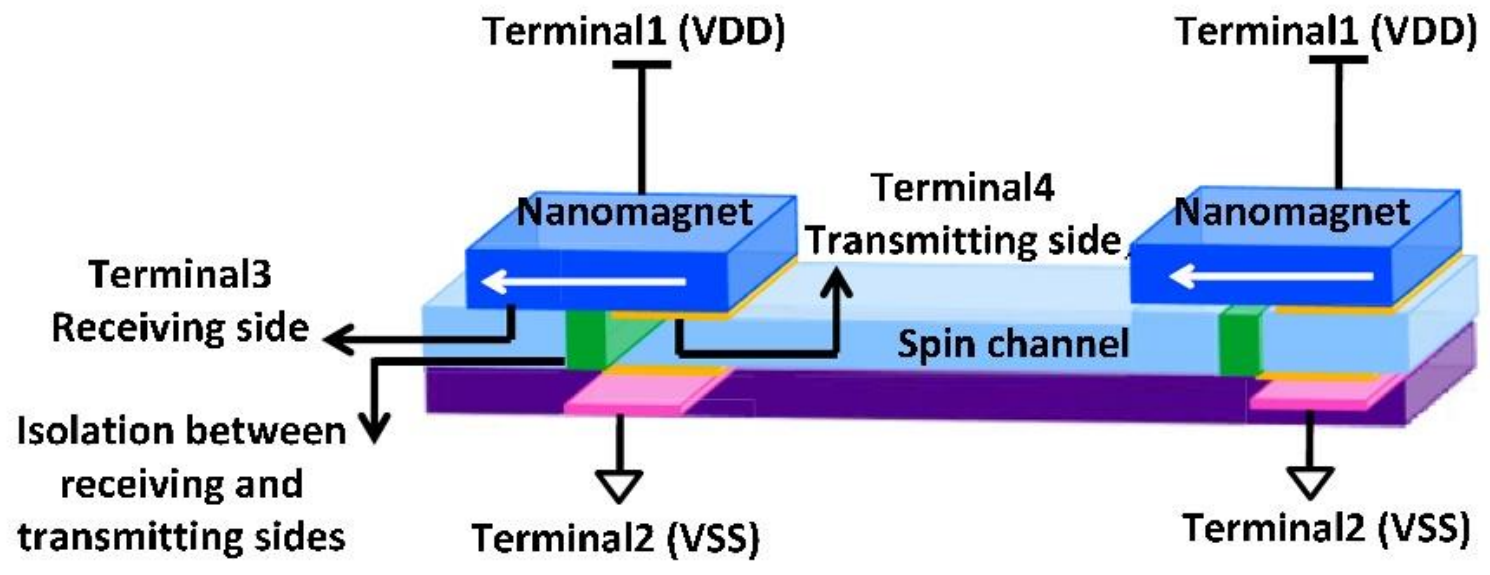

**b)**

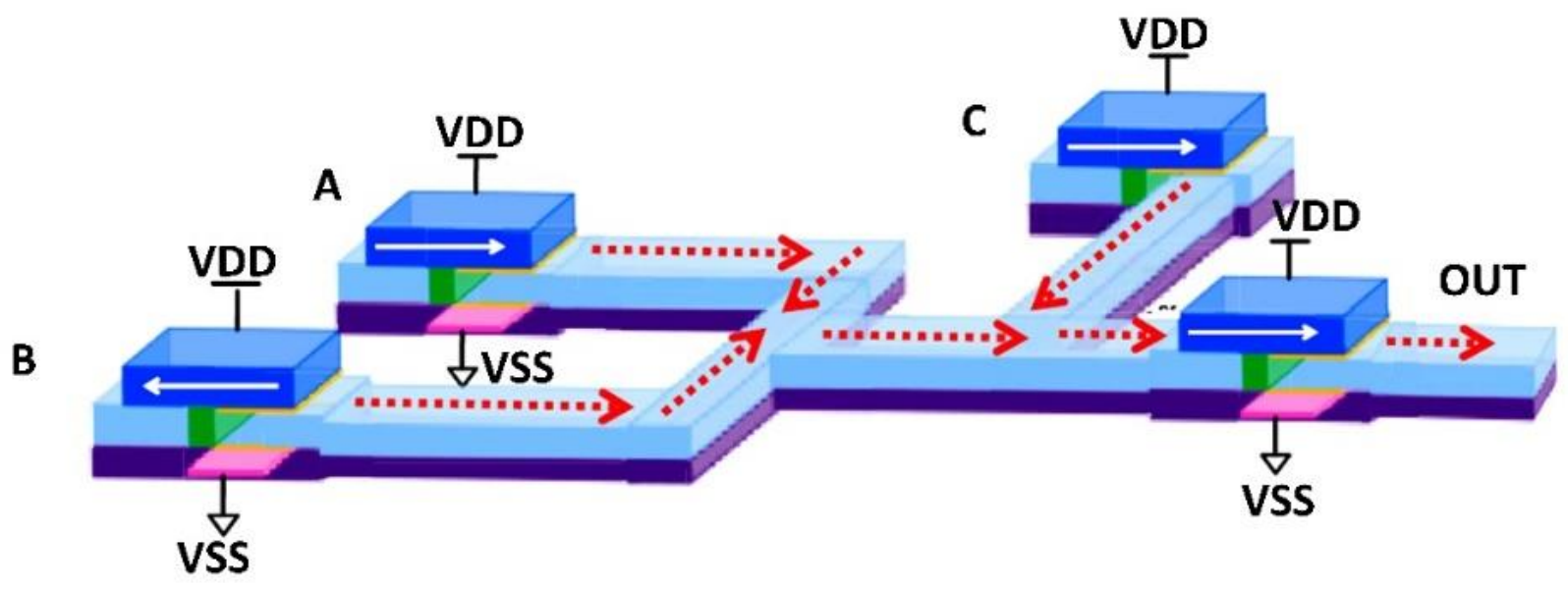

**c)**

Figure 3: a) Layout of an ASL device. b) Layout of inverter using ASL_NC. c) ASL majority gate M3 (direction of information propagation is shown with arrows) [6].

### 2.4 SET technology

Figure 4.a) shows a basic minority SET gate. The inputs with three capacitors form a voltage summing system which generates a mean voltage at node $A$. If this voltage exceeds a certain threshold, an electron tunnels through Single Electron Boxes (SEBs) and negates the voltage at $A$. Otherwise, the voltage remains positive. Logic 1 and logic 0, are represented by positive and negative voltages respectively.

A majority gate implemented by a balanced pair of single-electron boxes is shown in Figure 4.b) [4]. An electron tunnels through one of the SEBs to make a negative voltage and prevents movements of other electron when VDD increases. Hence, the stable voltage states for the two SEBs are (1, 0) and (0, 1) based on the inputs. For example, if all inputs are 0, the voltage state is (0, 1) and node $B$ has a negative voltage.

By fixing one of the three inputs to logic 0 or 1, “NAND” gate and “NOR” gate are achieved for SET minority gate, while “AND” gate and NOR gate are obtained for SET majority gate [3].

### 2.5 TPL Technology

Figure 4.c) shows a minority gate in TPL that uses two phases. The phase of a waveform is used in TPL to represent logic values in digital circuits. $C_j$ indicates the tunneling junction capacitance. The operation of TPL is based on the phase locking of SET oscillations to a pump signal that is distributed throughout the circuit. Since the pump frequency is set to twice the tunneling frequency, the electrical phase of the locked oscillation can have two different values. Each value represents a binary encoding [5], [32]. A TPL minority gate can be a two-input NOR or NAND gate by fixing one of its inputs to logic high or logic low, respectively.

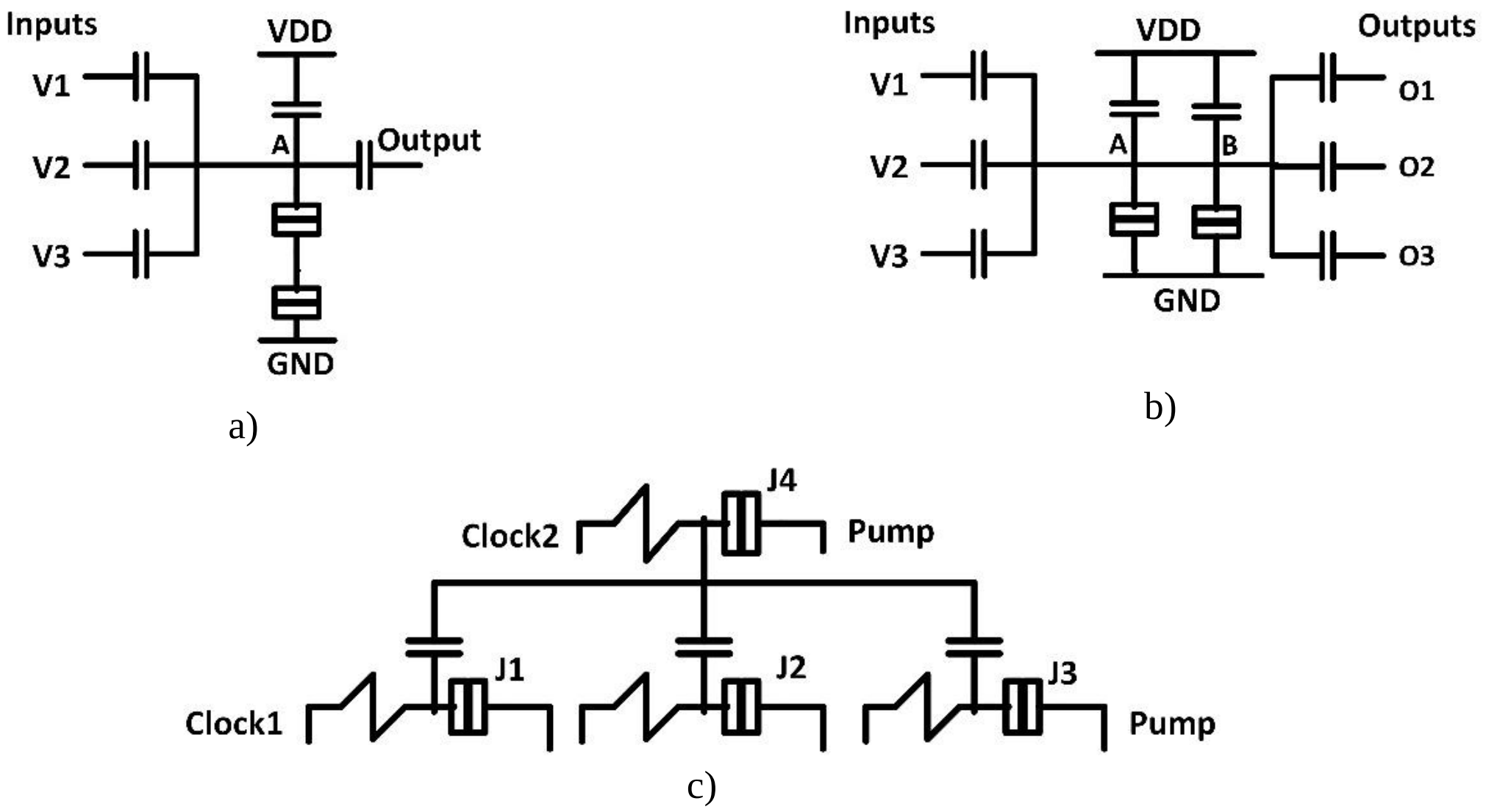

Figure 4: a) SET minority gate. b) SET majority gate. c) TPL minority gate

## 3 Proposed method

The suggested synthesis method is based on the creation of Majority Specification Matrix (MSM). In the mentioned matrix, all of the input states of a majority function are placed in columns of this matrix. In fact, the specification of the majority function output for each certain input state $(a, b, c)$ is placed in each of MSM columns. As there are $2^3$ possible input states for a three-input majority gate as shown in Figure 5, the dimensions of the matrix are 8 x 8, and it can be considered as a regular matrix. More details on this topic can be found in [33-35]. In this matrix, binary number of each column is related to a certain majority function, for instance, the binary number of the second column is $001$ which is equivalent to $Maj(a', b', c)$; it means that zeros in the binary number of each column are related to a inverter gate in the input of majority gate with a certain order.

The following features can be specified in the mentioned matrix:

- Specification of output of pair columns (4, 5), (3, 6), (2, 7), and (1, 8) are complementary.
- With respect to each of the two non-complementary columns, there are exactly two input states with the value of one, which are common in each of the two columns.
- If two majority gates are common in two input variables, then the following properties are held:

$$Maj(a,b,c) + Maj(a,b,c') = Maj(a,b,c+c') = Maj(a,b,1), \quad (3)$$
$$Maj(a,b,c) \times Maj(a,b,c') = Maj(a,b,c \times c') = Maj(a,b,0).$$

- Changing the order of input variables does not change the specification function in each column.

| Maj(a’,b’,c’) | Maj(a’,b’,c) | Maj(a’,b,c’) | Maj(a’,b,c) | Maj(a,b’,c’) | Maj(a,b’,c) | Maj(a,b,c’) | Maj(a,b,c) |
|---|---|---|---|---|---|---|---|
| 000 | 001 | 010 | 011 | 100 | 101 | 110 | 111 |
| 0 | 1 | 2 | 3 | 4 | 5 | 6 | 7 |

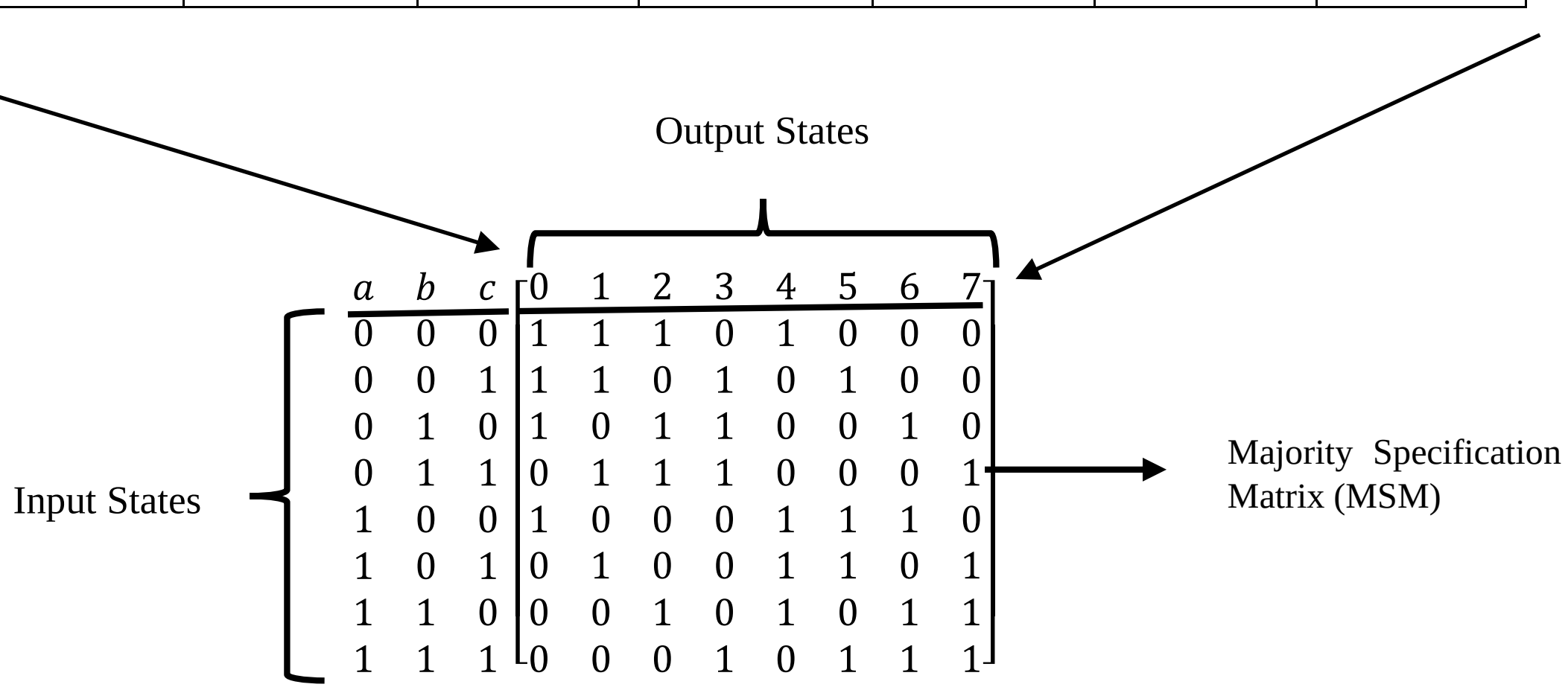

Figure 5: Structure representation of Majority Specification Matrix (MSM).

In the following sections, the application of MSM for synthesis of Boolean functions will be elaborated on.

### 3.1 Three-input Boolean functions

This section explains the application of MSM for the synthesis of three-input Boolean logic functions. To this end, the proposed methods are divided into two basic parts and a post-processing method. These methods have been generally designed to achieve the following two objectives:

- First, the simplest expression based on majority function should be achieved for each function
- The number of common expressions in the outputs of multi-output functions should be the maximum possible number and the minimum number of inverter gates should exist in them.

#### 3.1.1 Base of Method 1

In the first method, one majority gate and AND/OR functions are used. First, the specification function is compared to each column of MSM. One of the columns with the most identical number of ones is selected, i.e. Hamming code created between columns of MSM with the given specification function had the minimum possible number of ones. Then, through the application of AND and OR functions, additional ones are removed, and minterms with additional zeros are converted to ones, respectively. Following this step, for each part K-map is used for further simplification of the final expression. In Figure 6, an overall schematic of Method 1 is shown.

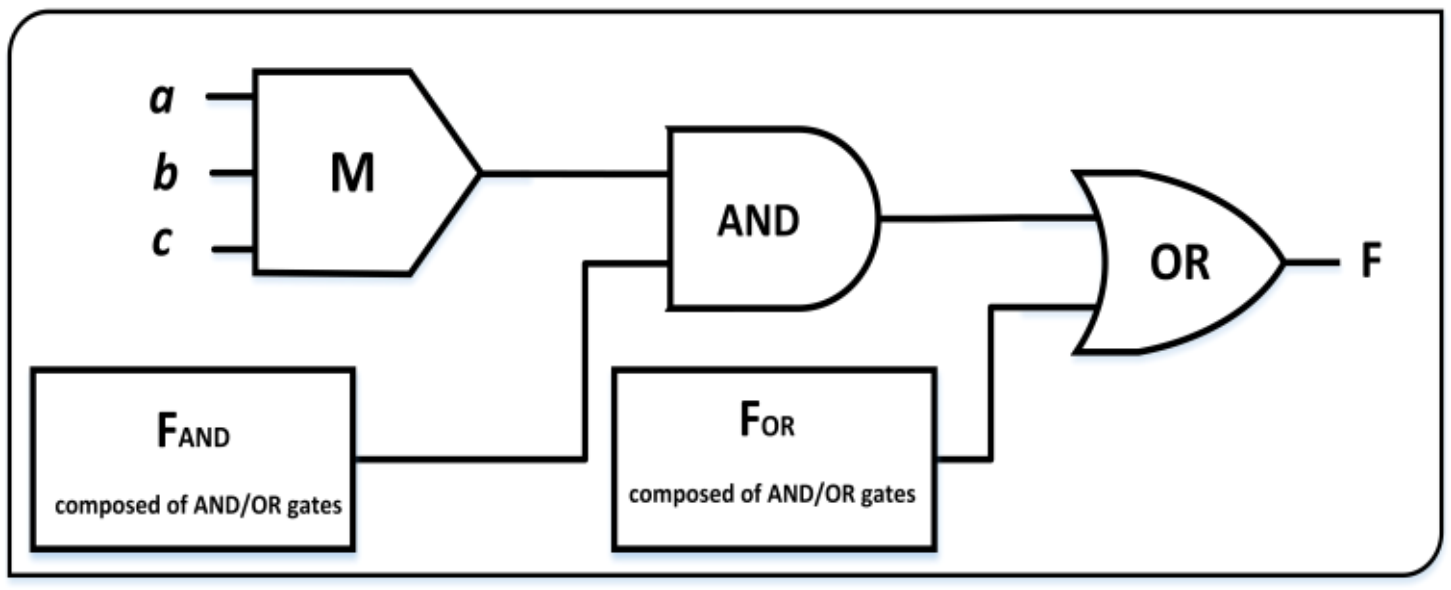

Figure 6: An overall schematic of Method 1.

The impact of AND and OR functions on the main function output and selected majority gate are presented in Table 1. In this table, the impact of AND gate applied between selected majority gate and $F_{And}$ then OR gate applied between majority gate and $F_{OR}$ are shown respectively. As AND and OR gates are complementary in their usage, the order of using them is not important. In Method 1, first AND and then OR gates are applied, respectively.

Table 1: The impact of applied AND/OR between pairs (selected majority gate, $F_{AND}$) then (applied majority gate, $F_{OR}$), respectively.

| Output function | Majority gate selected | $F_{AND}$ | $F_{OR}$ |
|---|---|---|---|
| 0 | 1 | 0 | 0 |
| 0 | 0 | $X$ | 0 |
| 1 | 0 | $X$ | 1 |
| 1 | 1 | 1 | $X$ |

In this table, $X$ denotes "don't care" state. Moreover, the majority expressions for AND and OR gates are as follows:

$$AND = Maj(f_1, f_2, 0), \qquad OR = Maj(f_1, f_2, 1), \tag{4}$$

In these equations, $f_1$ and $f_2$ functions are obtained from Method 1. The following example explains this method in more detail:

Example 1: Consider specification $F\ (a, b, c)\ =\ (0, 3, 6)$ (the numbers represent minterms contained in the function) defined as the first and the second columns of Table 2.

Table 2: Representation of the specification function related to Example 1 by the application of the proposed Method 1.

| $a$ | $b$ | $c$ | $F$ | $F(2) = M(a', b, c')$ | $F_{And}$ |
|---|---|---|---|---|---|
| 0 | 0 | 0 | 1 | 1 | 1 |
| 0 | 0 | 1 | 0 | 0 | $X$ |
| 0 | 1 | 0 | 0 | 1 | 0 |
| 0 | 1 | 1 | 1 | 1 | 1 |
| 1 | 0 | 0 | 0 | 0 | $X$ |
| 1 | 0 | 1 | 0 | 0 | $X$ |
| 1 | 1 | 0 | 1 | 1 | 1 |
| 1 | 1 | 1 | 0 | 0 | $X$ |

As shown in Table 2, Column 2 of MSM is selected as the most similar column to the specification function ($M(a', b, c')$). There is an additional minterm, in which function value is 1, i.e., 010. By the application of AND function as shown in Column 4 of Table 2, additional one value is converted to zero value. In Table 2, AND operation must be applied between Columns 3 and 4. It must be noted that Column 4 is created using Table 1. For simplification of the function in Column 4, K-map is used as shown in Table 3.

Table 3: Simplification of $F_{AND}$ created by K-map.

| $a$ \ $bc$ | 00 | 01 | 11 | 10 |
|---|---|---|---|---|
| 0 | 1 | $X = 1$ | 1 | 0 |
| 1 | $X = 1$ | $X = 1$ | $X = 1$ | 1 |

As presented in Table 3, for further simplification of function, all "don't care" states receive a value of one. Hence, one possible majority expression for $F_{AND}$ can be obtained as follows:

$$F_{AND} = a + b' + c = Maj(a, Maj(b', c, 1), 1). \quad (5)$$

The final result is provided as follows.

$$F = Maj(Maj(a', b, c'), F_{AND}, 0) = Maj(Maj(a', b, c'), Maj(a, Maj(b', c, 1), 1), 0). \quad (6)$$

3.1.1.1 Another structure of Method 1

It is worth mentioning that instead of specifying the most similar column from MSM, it is better to compare the columns of function input ($a, b, c$) with the specification of function output. It is due to this issue that in some functions, the places of ones in input columns are most similar to those of output columns. Hence, one of the majority gates would be removed. This method can also be used for the proposed methods in the next sections.

3.1.2 Base of Method 2

In the second method, a majority gate with three inputs $f_1, f_2, f_3$ is employed ($F = Maj(f_1, f_2, f_3)$); then, expressions related to the three functions are obtained. In Figure 7, an overall schematic of Method 2 is shown.

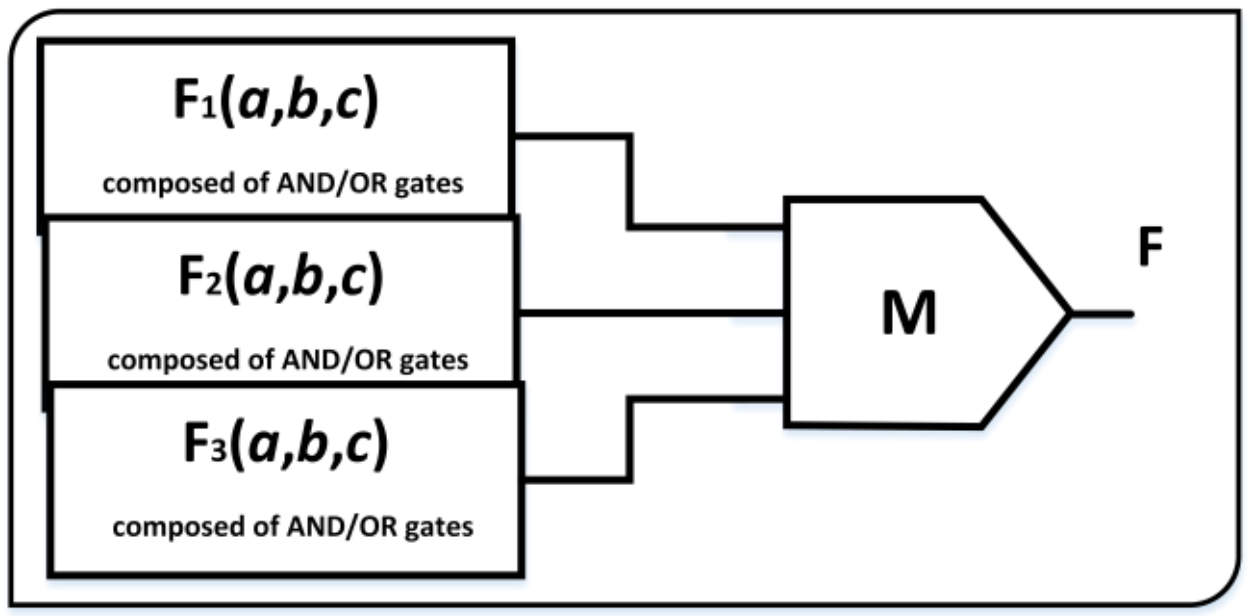

Figure 7: An overall schematic of Method 2.

In this method, similar to Method 1, first, the column of MSM which is the most similar to the function specification, is selected ($f_1$), and then the function of $f_2$ is obtained according to Table 4.

Table 4: Obtaining $f_2$ function according to $f_1$ and main function.

| Main Function | $f_1$= Majority Gate selected | $f_2$ |
|---|---|---|
| 0 | 0 | $X$ |
| 0 | 1 | 0 |
| 1 | 0 | 1 |
| 1 | 1 | $X$ |

As demonstrated in Table 4, in the process of comparing the main function and the selected majority gate, if the minterm has hamming code one, then the value of function $f_2$ must be equal to the value of main function; otherwise, the value of minterm in function $f_2$ should be "don't care" ($X$). The mentioned point is due to the feature of majority function. If the number of ones in a minterm is greater than or equal to two, then the value of output of majority function should be one; otherwise it should be zero. Then, for further simplification of Function $f_2$, K-map is used. When the value of "don't care" states is determined considering the above stated feature for majority gate, the value of minterms of function $f_3$ can be obtained considering Table 4. The following example illustrates the idea:

Example 2: Consider specification $F\ (a, b, c)\ =\ (1,2,4,5,6,7)$ defined in the second column of Table 5.

Table 5: Representation of specification function related to Example 2 after the application of Method 2.

| $a$ | $b$ | $c$ | $F$ | $f_1 = Maj(a', b', c')$ | $f_2$ | $f_3$ |
|---|---|---|---|---|---|---|
| 0 | 0 | 0 | 0 | 1 | 0 | 0 |
| 0 | 0 | 1 | 1 | 1 | $X = 0$ | 1 |
| 0 | 1 | 0 | 1 | 1 | $X = 0$ | 1 |
| 0 | 1 | 1 | 0 | 0 | $X = 0$ | $X = 1$ |
| 1 | 0 | 0 | 1 | 1 | $X = 1$ | $X = 0$ |
| 1 | 0 | 1 | 1 | 0 | 1 | 1 |
| 1 | 1 | 0 | 1 | 0 | 1 | 1 |
| 1 | 1 | 1 | 1 | 0 | 1 | 1 |

First, the column of MSM with the least difference with the main function is selected ($f_1 = Maj(a', b', c')$). Then, the value of function $f_2$ is obtained according to Table 4 and K-map presented in Table 6 as follows ($f_2 = a$).

Table 6: Representation of K-map applied for simplification of $f_2$ function.

| $bc$ \ $a$ | 00 | 01 | 11 | 10 |
|---|---|---|---|---|
| 0 | 0 | $X = 0$ | $X = 0$ | $X = 0$ |
| 1 | $X = 1$ | 1 | 1 | 1 |

Value of function $f_3$ is obtained by considering $f_2$ and Table 4; for example, the value obtained for $f_2$ in state (001) is zero, which is not equivalent to value $f_1$ and main function. Hence, value $f_3$ would be equal to the value of main function. Furthermore, the values of $f_2$, $f_1$, and the main function in state (011) are the same; therefore, it can be stated that the value of $f_3$ is "don't care". For further simplification of $f_3$, K-map is used as presented in Table 7.

Table 7: Applying K-map for simplification of $f_3$ function.

| $bc$ \ $a$ | 00 | 01 | 11 | 10 |
|---|---|---|---|---|
| 0 | 0 | 1 | $X = 1$ | 1 |
| 1 | $X = 0$ | 1 | 1 | 1 |

Logic expression for $f_3$ is as follows:

$$f_3 = (b + c) = Maj(b, c, 1) \tag{7}$$

The total logic expression for main function ($F$) is:

$$F = Maj(Maj(a', b', c'), a, Maj(b, c, 1)) \tag{8}$$

#### 3.1.2.1 Another structure of Method 2

For some functions, combination of the first (AND, OR) and the second methods can lead to better results. In the following example, the above-mentioned method is explained.

Example 3. Consider function $F\ (a,b,c)\ =\ (3,4,6)$ as shown in Table 8.

Table 8: An example of the combination of the first and the second methods for improvement of the results.

| $a$ | $b$ | $c$ | $F$ | $f_1 = Maj(7) = Maj(a,b,c)$ | $F_{OR}$ | $F_{tot}$ | $f_2$ | $f_3$ |
|---|---|---|---|---|---|---|---|---|
| 0 | 0 | 0 | 0 | 0 | 0 | 0 | $X=0$ | $X$ |
| 0 | 0 | 1 | 0 | 0 | 0 | 0 | $X=0$ | $X$ |
| 0 | 1 | 0 | 0 | 0 | 0 | 0 | $X=0$ | $X$ |
| 0 | 1 | 1 | 1 | 1 | 0 | 1 | $X=0$ | 1 |
| 1 | 0 | 0 | 1 | 0 | 1 | 1 | $X=0$ | 1 |
| 1 | 0 | 1 | 0 | 1 | 1 | 1 | 0 | 0 |
| 1 | 1 | 0 | 1 | 1 | 1 | 1 | $X=0$ | 1 |
| 1 | 1 | 1 | 0 | 1 | 1 | 1 | 0 | 0 |

First, the most similar column of MSM to the main function is selected ($f_1 = Maj(a,b,c)$). By selecting function $f_1$ as presented in Table 8 and the examination of K-map, it can be observed that in order to apply the OR operation between $f_1$ and $F_{OR}$ (Columns 3 and 4 in Table 8), all of the needed 1's states should be created without adding extra majority function. It means that the logic expression for $F_{OR}$ is ($F_{OR} = a$, as shown in Figure 8).

| $bc$ \ $a$ | 00 | 01 | 11 | 10 |
|---|---|---|---|---|
| 0 | 0 | 0 | 1 | 0 |
| 1 | 0 | 1 | 1 | 1 |

Adding OR →

| $bc$ \ $a$ | 00 | 01 | 11 | 10 |
|---|---|---|---|---|
| 0 | 0 | 0 | 1 | 0 |
| 1 | 1 | 1 | 1 | 1 |

F_OR=a

Figure 8: Applying OR operation to $f_1$ function.

The obtained result is shown in Column 5 ($F_{tot} =\ Maj(Maj(a,b,c),a,1)$). Hence, by consideration of Table 4, comparison of values related to $F_{tot}$ and the main function, and by application of K-map, the value of $f_2$ can be obtained. With respect to the specification function $f_2$ obtained in Column 6, it is certain that the logic function $f_2$ is zero ($f_2 = 0$) as it is composed of "don't care" and "zero" states. Hence, the simplest function with the minimum number of majority gates is the zero function. Moreover, the specification function $f_3$ is observed to be the same as $f_2$. Then, the logic function is simplified by applying K-map (as shown in Table 9). Thus, the logic expression for $f_3$ is:

$$f_3 = a' + c' = Maj(a',c',1) \quad (9)$$

Table 9: Applying K-map for simplification of $f_3$ function.

| $bc$ \ $a$ | 00 | 01 | 11 | 10 |
|---|---|---|---|---|
| 0 | $X=1$ | $X=1$ | 1 | $X=1$ |
| 1 | 1 | 0 | 0 | 1 |

As a result, the total logic expression obtained for main function ($F$) is:

$$F = Maj(Maj(Maj(a, b, c), a, 1), 0, Maj(a', c', 1)) \tag{10}$$

### 3.1.3 Post-processing method

To obtain better results, the post-processing method is used, which can be applied to the above presented methods and can improve the obtained results. In this method, after applying Method 2 to the specification function and obtaining functions of $f_2$ and $f_3$, K-maps related to $f_2$ and $f_3$ are studied simultaneously. It is conjectured that if some of 1's states related to $f_2$ and $f_3$ are exchanged, better results would be obtained. The mentioned point is due to the feature of majority function. If the number of ones in a minterm is greater than or equal to two, then the value of output of majority function would be one; otherwise it would be zero. Furthermore, the following steps are taken into account in this method:

- In the K-maps of $f_2$ and $f_3$, minterms of main function whose corresponding function value is one are marked. Due to the mentioned feature for the majority gate, these places (cubes of K-map) can have numbers of 2 or 3 ones in the majority function (Group 1).
- In the K-maps of $f_2$ and $f_3$, minterms of main function with the zero value of function can have numbers of 0 or 1 ones in the majority function (Group 2).
- Fixed minterms in the K-map $f_2$ are not considered.
- "Don't care" states in $f_2$, which do not create a square in the K-map and generate an extra state in the K-map related to $f_3$, are exchanged in response to the rules of Groups 1 and 2. This method as a search method is continued until the best square in the K-maps is obtained.
- Equation (11) holds for each majority gate. By applying that expression, the number of inverter gates in the general expression can be decreased.

$$Maj(a', b', c') = Maj(a, b, c)' \tag{11}$$

Generally, only states of $f_2$, which are "don't care", are considered. Then, the states generated in $f_3$ which do not make a square in $f_2$ are taken into account, and these states are exchanged between $f_2$ and $f_3$ to make squares in $f_2$ and $f_3$. Then, the squares providing the least number of majority expressions are selected from the overall obtained squares.

- Proposition: Suppose $f, g, u, d$ are Boolean functions and the following expression exists between them: $y = u(fg' + gd)$, where $g'$ is the complement of $g$. Then, the equivalent majority expression is as the following expression: $y = M(u, M(f, g', 0), M(g, d, 0))$

  Prove: By extending equivalent majority expression, we have:

$$y = u(fg' + gd) + \cancelto{0}{\big((fg') \times (gd)\big)} = u(fg' + gd).$$

## 3.2 Four and higher-input Boolean functions

In this section, for the synthesis of four- and higher-input Boolean functions, it must be considered that the majority gate has three inputs; thus, for instance, the methods are explained for four-input functions. Accordingly, the methods presented in this section can be generalized to higher inputs.

Taking into account that the majority gate has three inputs; thus, for four-input functions with consideration of the inputs intended to function, there would be four permutations ($C(3,4) = 4$, where, $C$ denotes the combination of permutations.), as the order of inputs does not matter. It is assumed that inputs of the main function are $a, b, c, and, d$. Moreover, it must be mentioned that input (a) is the most significant one, and input (d) is the least significant one.

$$Maj(b, c, d)\ , Maj(a, b, c), Maj(a, c, d), Maj(a, b, d) \tag{12}$$

For each of the above-mentioned permutations, an MSM must be created according to Figure 5. For the creation of MSM with $b, c, and\ d$ inputs, two matrices of MSM can be created under each other. It means that two matrices of MSM are placed in one column as cascades. As a result, the matrix of MSM has 16 rows and 8 columns. For the creation of MSM in the other states, each row of MSM created in Figure 5 must move to two new rows for four-input functions. For example, consider the $Maj(a, b, c)$ state. For the creation of the mentioned state, the majority function $Maj(a, b, c)$ must shift variables $(a, b, c)$ to the left in comparison to state $Maj(b, c, d)$. In Table 10, this method is shown. In fact, variable $d$ in the main permutation $(b, c, d)$ is as a "don't care" variable for new permutation $(a, b, c)$.

Table 10: Creation of new states in four-input functions according to the three-input functions.

| Old rows | New permutation of rows | Old rows | New permutation of rows |
|---|---|---|---|
| $b\ c\ d$ | $a\ b\ c\ \boldsymbol{d}$ | $b\ c\ d$ | $a\ b\ c\ \boldsymbol{d}$ |
| 0 0 0 | 0 0 0 **0**<br>0 0 0 **1** | 1 0 0 | 1 0 0 **0**<br>1 0 0 **1** |
| 0 0 1 | 0 0 1 **0**<br>0 0 1 **1** | 1 0 1 | 1 0 1 **0**<br>1 0 1 **1** |
| 0 1 0 | 0 1 0 **0**<br>0 1 0 **1** | 1 1 0 | 1 1 0 **0**<br>1 1 0 **1** |
| 0 1 1 | 0 1 1 **0**<br>0 1 1 **1** | 1 1 1 | 1 1 1 **0**<br>1 1 1 **1** |

For four-input functions, the following methods are used:

1. In this method, the majority gate is used as a tree expression and employs Method 2 discussed in the three-input functions section. In the mentioned section, if the order of ones in each $f_2$ or $f_3$ functions leads to the creation of complex specification functions, as a result many majority gates will be created in each function. Hence, for the synthesis of $f_2$ or $f_3$ functions, each of them can be considered as a main function. Then, in the process of applying Method 2 explained in three-input functions section, the mentioned functions can be synthesized once more. This method can be repeatedly carried out to determine its acceptable level. In Figure 9, an overall schematic of this method is shown.

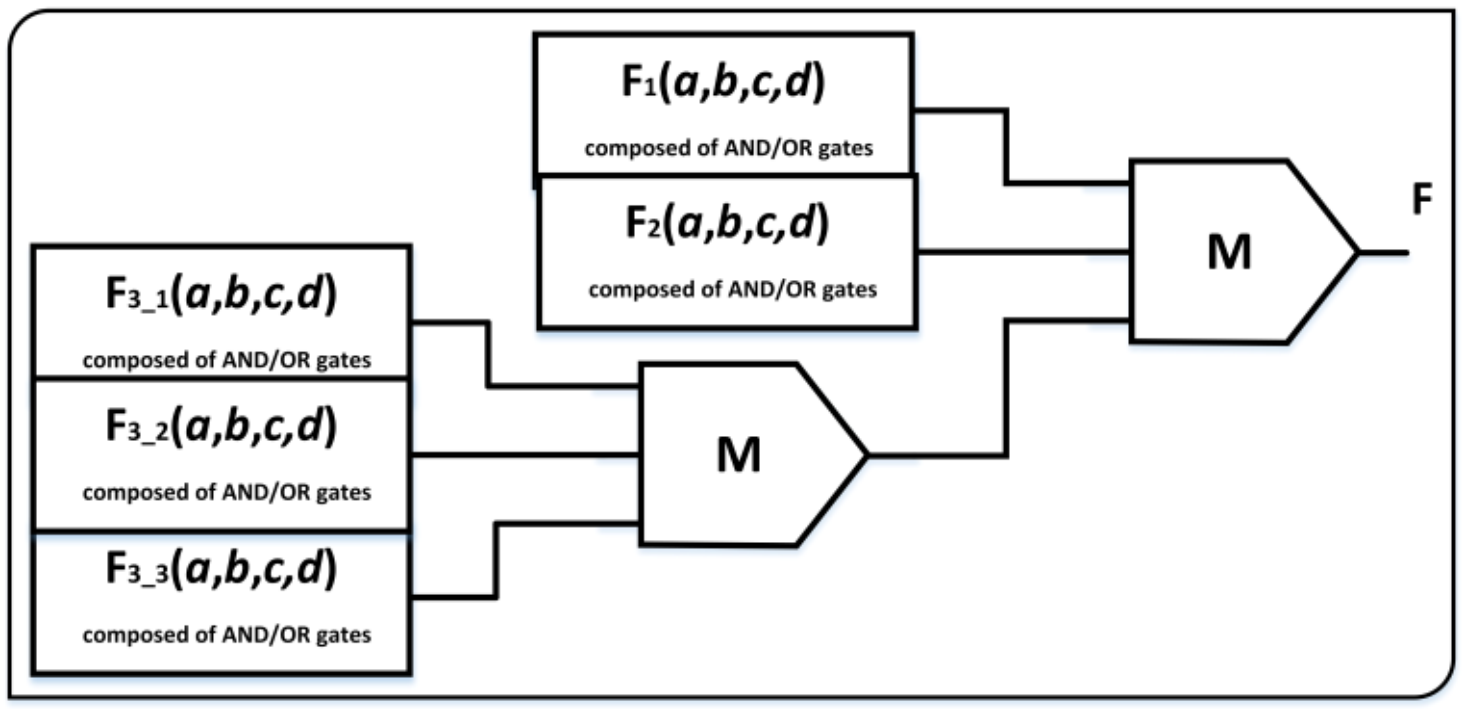

Figure 9: An overall schematic of Method 3.

2. For synthesis of functions larger than three-inputs, the methods explained in the previous section can be used.

In the following examples, the above-mentioned methods are explained.

Example 4. Consider function $F = (9,11,14)$ defined as the first and the second columns of Table 11.

Table 11: Specification function used in Example 4 and calculation of $f_2$ and $f_3$ functions.

| $a$ | $b$ | $c$ | $d$ | $F$ | $f_1 = M(7) = M(a,b,d)$ | $f_2$ | $f_3$ |
|---|---|---|---|---|---|---|---|
| 0 | 0 | 0 | 0 | 0 | 0 | $X=0$ | $X=1$ |
| 0 | 0 | 0 | 1 | 0 | 0 | $X=0$ | $X=1$ |
| 0 | 0 | 1 | 0 | 0 | 0 | $X=0$ | $X=1$ |
| 0 | 0 | 1 | 1 | 0 | 0 | $X=0$ | $X=1$ |
| 0 | 1 | 0 | 0 | 0 | 0 | $X=0$ | $X=0$ |
| 0 | 1 | 0 | 1 | 0 | 1 | 0 | 0 |
| 0 | 1 | 1 | 0 | 0 | 0 | $X=0$ | $X=1$ |
| 0 | 1 | 1 | 1 | 0 | 1 | 0 | 0 |
| 1 | 0 | 0 | 0 | 0 | 0 | $X=0$ | $X=1$ |
| 1 | 0 | 0 | 1 | 1 | 1 | $X=0$ | 1 |
| 1 | 0 | 1 | 0 | 0 | 0 | $X=0$ | $X=1$ |
| 1 | 0 | 1 | 1 | 1 | 1 | $X=0$ | 1 |
| 1 | 1 | 0 | 0 | 0 | 1 | 0 | 0 |
| 1 | 1 | 0 | 1 | 0 | 1 | 0 | 0 |
| 1 | 1 | 1 | 0 | 1 | 1 | $X=0$ | 1 |
| 1 | 1 | 1 | 1 | 0 | 1 | 0 | 0 |

According to Method 2 explained in three-input functions, first, the column of MSM that is the most similar one to the specification function is selected ($f_1 = M(a,b,d)$). Then, specification functions ($f_2$ and $f_3$) are obtained. It can be certainly stated that logic function $f_2$ is zero ($f_2 = 0$), and the logic function $f_3$ is obtained by the application of K-map as shown in Table 12:

Table 12: Applying K-map for obtaining $f_3$ function.

| $ab$ \ $cd$ | 00 | 01 | 11 | 10 |
|---|---|---|---|---|
| 00 | $X=1$ | $X=1$ | $X=1$ | $X=1$ |
| 01 | $X=0$ | 0 | 0 | $X=1$ |
| 11 | 0 | 0 | 0 | 1 |
| 10 | $X=1$ | 1 | 1 | $X=1$ |

Boolean logic function $f_3$ is:

$$f_3 = b' + cd' = Maj(b', Maj(c, d', 0), 1) \quad (13)$$

By the implementation of the post-processing method presented in three-input functions section, specification functions of $f_2$ and $f_3$ can be exchanged as shown in Figure 10. States of main specification function, which have the value of one, are presented by gray color in Figure 10. Moreover, values of states 0000 and 1000 have been changed to zero in Figure 10(b). As the mentioned states in the main specification function are zero, the number of ones placed in these cubes can be zero or one. In this example, Column 10 is common to two Rows 00, 10, which have been exchanged with the same column in Figure 10(a) ($f_2$).

| $ab \backslash cd$ | 00 | 01 | 11 | 10 |
|---|---|---|---|---|
| 00 | $X=0$ | $X=0$ | $X=0$ | $X=0$ |
| 01 | $X=0$ | 0 | 0 | $X=0$ |
| 11 | 0 | 0 | 0 | $X=0$ |
| 10 | $X=0$ | $X=0$ | $X=0$ | $X=0$ |

Rows 00, 01, 11, 10 (column 10) → $X = 1$

(a)

| $ab \backslash cd$ | 00 | 01 | 11 | 10 |
|---|---|---|---|---|
| 00 | $X=1$ | $X=1$ | $X=1$ | $X=1$ |
| 01 | $X=0$ | 0 | 0 | $X=1$ |
| 11 | 0 | 0 | 0 | 1 |
| 10 | $X=1$ | 1 | 1 | $X=1$ |

Row 00, column 00 → X=0; row 10, column 00 → $X = 0$; rows 00, 01, 11, 10 (column 10) → $X = 0$

(b)

Figure 10: (a) Applying K-map for $f_2$ function, (b) Applying K-map for $f_3$ function-Applying post-processing method to $f_2$ and $f_3$.

Then, states (1001, 1011) are fixed and thus, among of other states, states (0000, 0100) must be zero until it creates a square in K-map. Logic expressions for $f_2$ and $f_3$ are:

$$f_2 = cd' = Maj(c, d', 0),\ \ f_3 = b'd = Maj(b', d, 0). \tag{14}$$

The total logic function is:

$$F = Maj(Maj(a, b, d), Maj(c, d', 0), Maj(b', d, 0)) \tag{15}$$

Example 5. Consider the specification function $F = (3,4,7,15)$ is defined as the first and the second columns of Table 13. The tree method is explained for this example.

Table 13: Specification function used in Exp. 5 and calculation of $f_2$ and $f_3$.

| $A$ | $B$ | $C$ | $D$ | $F$ | $f_1 = M(0, c, d)$ | $f_2$ | $f_3$ |
|---|---|---|---|---|---|---|---|
| 0 | 0 | 0 | 0 | 0 | 0 | $X=0$ | $X$ |
| 0 | 0 | 0 | 1 | 0 | 0 | $X=0$ | $X$ |
| 0 | 0 | 1 | 0 | 0 | 0 | $X=0$ | $X$ |
| 0 | 0 | 1 | 1 | 1 | 1 | $X=0$ | 1 |
| 0 | 1 | 0 | 0 | 1 | 0 | 1 | 1 |
| 0 | 1 | 0 | 1 | 0 | 0 | $X=1$ | 0 |
| 0 | 1 | 1 | 0 | 0 | 0 | $X=1$ | 0 |
| 0 | 1 | 1 | 1 | 1 | 1 | $X=1$ | $X$ |
| 1 | 0 | 0 | 0 | 0 | 0 | $X=0$ | $X$ |
| 1 | 0 | 0 | 1 | 0 | 0 | $X=0$ | $X$ |
| 1 | 0 | 1 | 0 | 0 | 0 | $X=0$ | $X$ |
| 1 | 0 | 1 | 1 | 0 | 1 | 0 | 0 |
| 1 | 1 | 0 | 0 | 0 | 0 | $X=1$ | 0 |
| 1 | 1 | 0 | 1 | 0 | 0 | $X=1$ | 0 |
| 1 | 1 | 1 | 0 | 0 | 0 | $X=1$ | 0 |
| 1 | 1 | 1 | 1 | 1 | 1 | $X=1$ | $X$ |

First, the most similar column to main function is selected; that here, $Maj(0, c, d)$ is combined with $Maj(b, c, d)$ and $Maj(b', c, d)$. Then, the logic function $f_2$ according to K-map shown in Table 14 is obtained. ($f_2 = b$)

Table 14: K-map used for calculation $f_2$

| $ab$ \ $cd$ | 00 | 01 | 11 | 10 |
|---|---|---|---|---|
| 00 | $X=0$ | $X=0$ | $X=0$ | $X=0$ |
| 01 | 1 | $X=1$ | $X=1$ | $X=1$ |
| 11 | $X=1$ | $X=1$ | $X=1$ | $X=1$ |
| 10 | $X=0$ | $X=0$ | 0 | $X=0$ |

For obtaining the specification function $f_3$ as a tree method, its value is considered as the main function and is shown in Table 15.

Table 15: Consider $f_2$ function as main function for using to tree method.

| $a$ | $b$ | $c$ | $d$ | $f_3$ | $f_{3_1} = M(a', b', c')$ | $f_{3_2}$ | $f_{3_3}$ |
|---|---|---|---|---|---|---|---|
| 0 | 0 | 0 | 0 | $X$ | 1 | $X=0$ | $X=1$ |
| 0 | 0 | 0 | 1 | $X$ | 1 | $X=0$ | $X=0$ |
| 0 | 0 | 1 | 0 | $X$ | 1 | $X=0$ | $X=1$ |
| 0 | 0 | 1 | 1 | 1 | 1 | $X=0$ | 1 |
| 0 | 1 | 0 | 0 | 1 | 1 | $X=0$ | 1 |
| 0 | 1 | 0 | 1 | 0 | 1 | 0 | 0 |
| 0 | 1 | 1 | 0 | 0 | 0 | $X=0$ | $X=1$ |
| 0 | 1 | 1 | 1 | $X$ | 0 | $X=0$ | $X=1$ |
| 1 | 0 | 0 | 0 | $X$ | 1 | $X=0$ | $X=1$ |
| 1 | 0 | 0 | 1 | $X$ | 1 | $X=0$ | $X=0$ |
| 1 | 0 | 1 | 0 | $X$ | 0 | $X=0$ | $X=1$ |
| 1 | 0 | 1 | 1 | 0 | 0 | $X=0$ | $X=1$ |
| 1 | 1 | 0 | 0 | 0 | 0 | $X=0$ | $X=1$ |
| 1 | 1 | 0 | 1 | 0 | 0 | $X=0$ | $X=0$ |
| 1 | 1 | 1 | 0 | 0 | 0 | $X=0$ | $X=1$ |
| 1 | 1 | 1 | 1 | $X$ | 0 | $X=0$ | $X=1$ |

As shown in Table 15, function $f_3$ as main function is considered. It is important to note that the states of ”don’t care” in Function $f_3$ in calculation of specification Functions of $f_{3_1}$, $F_{3_2}$ and $f_{3_3}$ are do not care and it does not matter that their values be zero or one. First, the most similar column to the main function is selected ($M(a', b', c')$); and logic functions of $f_{3_2}$ and $f_{3_3}$ are:

$$f_{3_2} = 0, \quad f_{3_3} = d' + c = Maj(d', c, 1). \tag{16}$$

Then, specification functions of $f_{3_2}$ and $f_{3_3}$ are obtained by using the post-processing method as shown in Figure 11.

| cd / ab | 00 | 01 | 11 | 10 |
|---|---|---|---|---|
| 00 | $X = 0$ | $X = 0$ | $X = 0$ | $X = 0$ |
| 01 | $X = 0$ | 0 | $X = 0$ | $X = 0$ |
| 11 | $X = 0$ | $X = 0$ | $X = 0$ | $X = 0$ |
| 10 | $X = 0$ | $X = 0$ | $X = 0$ | $X = 0$ |

X=1

(a)

| cd / ab | 00 | 01 | 11 | 10 |
|---|---|---|---|---|
| 00 | $X = 1$ | $X = 0$ | 1 | $X = 1$ |
| 01 | 1 | 0 | $X = 1$ | $X = 1$ |
| 11 | $X = 1$ | $X = 0$ | $X = 1$ | $X = 1$ |
| 10 | $X = 1$ | $X = 0$ | $X = 1$ | $X = 1$ |

$X = 0$

(b)

Figure 11: (a) K-map used for function $f_{3_2}$, (b) K-map used for function $f_{3_3}$. Applying post-processing method to $f_{3_2}$and $f_{3_3}$ for more simplification.

New Boolean logic functions $f_{3_2}$ and $f_{3_3}$ are:

$$f_{3_2} = b', \qquad f_{3_3} = d'. \tag{17}$$

The total specification function is:

$$F = Maj(Maj(0, c, d), b, Maj(M(a', b', c'), b', d')). \tag{18}$$

## 4 A synthesis flow for multi-output functions

Having the proposed approaches from the previous section available, they can be combined to an extended synthesis flow that could be used for multi-output functions. Figure 12 illustrates this flow. As shown in this figure, first, according to the numbers of inputs of main function, MSMs are created, then for each of the outputs of function ($fi$) and their complementaries, the most similar column of MSM or inputs of function is selected; as well as, it could be selected from combination of columns in MSM that led to AND/OR functions. Then, for each of selected columns, the proposed methods in previous sections are applied. Afterwards, the obtained results are saved. In addition, conventional K-map method is applied separately and its result is saved. Because of that, in some functions, expression obtained from this method is simpler. Then, among of results obtained for each of outputs, for reaching to the most common expressions, in Line 10 of the algorithm, from expressions obtained in outputs for synthesis, the other outputs are used again. Then results based on the objectives of priority gate counts and gate levels are ordered, and after that results to the most common terms are selected. Finally, for reducing of the number of inverter gates in final expressions, Line 12 of algorithm is applied.

Input: a given specification of multi-output function
Output: a synthesized circuit to majority gate

1. Create the MSM according to the number of inputs of function
2. For each $fi$ ($i = 1{:}m$ $m$: the number of outputs of function)
3. { For each MSMs created
4. { Find the most similar column to specification of main function in MSM
5. Find the most similar column according to features of combination of columns in MSM.
6. Find the most similar specification between main function and columns of inputs.
7. Repeat Lines 4 and 5 for complementary function ($f_i$').
8. Apply all of the methods explained in previous sections to Lines 4 and 5 of algorithm and save all results.

9. Apply method of conventional K-map to main function and save the result.
}
}.
10. Use the majority expression created in each of $fi$ for synthesis other $fj$ ($j \neq i$)
11. Select results to the most common expressions between outputs.
12. For reducing the number of inverter gate, use the following feature $Maj(a', b', c')' = Maj(a, b, c)$

Figure 12: The proposed flow for the synthesis of multi-output functions.

## 5 Results and Comparison

In this section, first, a comparison between the proposed synthesis flow and a multi-objective genetic programming approach [24] is performed. The results are shown in Table 16. In this table, NOI, NOM and NTG stand for the number of inverter gates, number of majority gates and the total number of gates, respectively. The obtained synthesis results are shown in the last column of this table. In this column, the common parts of the corresponding synthesized circuit for each output are underlined. Also, in this table, the rows titled “shared gates” and “total number of gates” show the number of shared gates and the total number of gates used in the function outputs, respectively. The results illustrate that the proposed method produces the same or better results than the approach presented in [24].

In some functions, the two parameters of the number of gates and the number of common parts should be considered simultaneously; e.g., for the specification function $F = (2,6,10,11,14)$ the following expressions are obtained (this function is a sample of a multi-output function in Table 16):

$$F_1 = Maj(Maj(Maj(a, b', 0), d', 1), c, 0) \qquad (19)$$
$$F_2 = Maj(Maj(Maj(1', c, d), b, Maj(Maj(a, b, c), b, d)')', 1, c)$$

Although $F_1$ produces a fewer number of gates, $F_2$ is chosen because it leads to the more number of common parts with the other outputs presented in Table 16. The common parts lead to a more reduction in the number of total gates.

In Table 17, an overall comparison between the best existing majority logic synthesis method [29] and the proposed synthesis flow is demonstrated. Also, in this table the proposed synthesis flow is compared with the method presented in [28]. In this table, the first column lists the names of benchmarks. The columns titled “Method [28]” and “Method [29]” show the results for the corresponding benchmarks obtained from [28] and [29] in terms of the number of levels, gates and the number of inverter gates only for [29], as the method in [28] has not reported the number of inverter gates. The column titled “Proposed method” shows the results obtained from the proposed synthesis flow in terms of the number of levels, gates and the number of inverter gates. The “Reduction %” columns compare the proposed method with the methods in [28] and [29] and give the percentage reductions. This table illustrates that there is an average reduction of 14.9% in the number of levels and at the same time, the number of gates is reduced by 31.6% as compared to the method in [28]. When compared to [29], the average reduction in levels is 10.5% and the reduction in gate counts is 16.8% and also the reduction in inverter gates is 33.5%. Results show that the proposed synthesis flow outperforms the best existing methods. Also, the detailed results of the proposed synthesis flow applied to MCNC benchmarks are shown in appendix (Table A1). Since the synthesis results are independent of technology used, they are effective for any majority/minority-based technology including ASL, QCA, SET and TPL.

## 6 Conclusion

In this paper, a multi-objective synthesis methodology for generating optimal majority expressions was presented. In this method by using a Majority Specification Matrix (MSM) and a synthesis flow presented for multi-output specification functions, the majority expressions are optimized. The corresponding minority network can be easily obtained by complementing the majority expressions. The proposed approach was applied to 20 MCNC benchmarks and was compared with previous methods. Experimental results demonstrated the proposed method outperforms the best existing ones.

| Function [24] | [24] | | | The Proposed Method | | | The Synthesized Circuit |
|---|---|---|---|---|---|---|---|
| | NOI | NOM | NTG | NOI | NOM | NTG | |
| $f_1$ = (2,4,6) | 3 | 4 | 7 | 1 | 2 | 3 | M(M(a,b,c'),0,c') |
| $f_2$ = (0,1,3,6) | 3 | 4 | 7 | 3 | 4 | 7 | M(M(a,M(a,b,c'),1)',1,M(a,b,1')) |
| $f_3$ = (0,3,6) | 4 | 4 | 8 | 4 | 4 | 8 | M(M(a',b,c'),M(M(a,b',1),c,1),1') |
| shared gates | 4 | 4 | 8 | 3 | 2 | 5 | |
| total number of gates | 6 | 8 | 14 | 5 | 8 | 13 | |
| $f_1$ = (1,4,5,7) | 1 | 1 | 2 | 1 | 1 | 2 | M(a,b',c) |
| $f_2$ = (3,4,6) | 3 | 4 | 7 | 3 | 4 | 7 | M(M(a,b,c),M(a',c',1),M(a,c',1')) |
| $f_3$ = (0,2,5,6) | 3 | 4 | 7 | 4 | 4 | 8 | M(M(a,b',c)',M(a,c',1),M(c,b',1')) |
| $f_4$ = (4,6,7) | 3 | 3 | 6 | 2 | 2 | 4 | M(M(a,c',1')),a,b) |
| shared gates | 4 | 3 | 7 | 5 | 2 | 7 | |
| total number of gates | 6 | 9 | 15 | 5 | 9 | 14 | |
| $f_1$ = (0,3,6,7,15) | 3 | 5 | 8 | 4 | 5 | 9 | M(M(a',b,M(c',d,1)),M(c,M(b,d,1)',1),1') |
| $f_2$ = (9,11,14) | 4 | 4 | 8 | 4 | 4 | 6 | M(M(a,b,d),1',M(a',b,M(c',d,1))') |
| $f_3$ = (8,10,11,14,15) | 1 | 3 | 4 | 2 | 3 | 5 | M(M(M(b,d,1)',c,1),a,1') |
| shared gates | 3 | 4 | 7 | 5 | 4 | 9 | |
| total number of gates | 5 | 8 | 13 | 5 | 8 | 13 | |
| $f_1$ = (3,4,7,15) | 3 | 4 | 7 | 2 | 4 | 6 | M(M(1',c,d),b,M(M(a,b,c),b,d)') |
| $f_2$ = (1,3,4,9,13,15) | 4 | 5 | 9 | 3 | 5 | 8 | M(M(M(a,b',1),d,1'),b,M(M(a,b,c),b,d)') |
| $f_3$ = (3,6,7,11,13,14,15) | 2 | 3 | 5 | 2 | 3 | 5 | M(M(M(a,b',1),d,1'),b,c) |
| $f_4$ = (2,6,10,11,14) | 5 | 5 | 10 | 3 | 5 | 8 | M(M(M(1',c,d),b,M(M(a,b,c),b,d)')',1,c) |
| shared gates | 6 | 8 | 14 | 6 | 8 | 14 | |
| total number of gates | 8 | 9 | 17 | 4 | 9 | 13 | |

Table 17: Comparisons of the proposed method with methods presented in [28] and [29].

| Benchmarks | Method [28] | | Method [29] | | | Proposed method | | | Reduction% VS. [28] | | Reduction% VS. [29] | | |
|---|---|---|---|---|---|---|---|---|---|---|---|---|---|
| | Level | Majority Gates | Level | Majority Gates | Inverters | Level | Majority Gates | Inverter | Level | Majority Gates | Level | Majority Gates | Inverter |
| b1 | 3 | 9 | 2 | 7 | 5 | 2 | 6 | 4 | 33.3% | 33.3% | 0.0% | 14.3% | 20.0% |
| cm82a | 4 | 16 | 3 | 7 | 6 | 3 | 6 | 4 | 25.0% | 62.5% | 0.0% | 14.3% | 33.3% |
| majority | 4 | 6 | 4 | 6 | 0 | 4 | 5 | 0 | 0.0% | 16.6% | 0.0% | 16.6% | 0.0% |
| 9symml | 12 | 216 | 10 | 47 | 23 | 10 | 47 | 18 | 16.6% | 78.2% | 0.0% | 0.0% | 21.7% |
| x2 | 7 | 42 | 7 | 37 | 15 | 6 | 34 | 11 | 14.2% | 19.0% | 14.3% | 8.1% | 26.6% |
| cm152a | 6 | 21 | 6 | 21 | 7 | 4 | 15 | 3 | 33.3% | 28.5% | 33.3% | 28.5% | 57.1% |
| cm85a | 7 | 34 | 6 | 26 | 12 | 6 | 14 | 9 | 14.3% | 58.8% | 0.0% | 46.1% | 25.0% |
| cm151a | 7 | 42 | 7 | 23 | 10 | 4 | 15 | 5 | 42.8% | 64.2% | 42.8% | 34.7% | 50.0% |
| cm162a | 7 | 46 | 7 | 41 | 14 | 8 | 32 | 11 | -14.3% | 30.4% | -14.3% | 21.9% | 21.4% |
| cu | 7 | 46 | 7 | 40 | 21 | 5 | 36 | 12 | 28.5% | 21.7% | 28.5% | 10.0% | 42.8% |
| cm163a | 7 | 42 | 7 | 38 | 17 | 6 | 28 | 16 | 14.3% | 33.3% | 14.3% | 26.3% | 5.0% |
| cmb | 4 | 44 | 4 | 28 | 4 | 4 | 26 | 2 | 0.0% | 40.9% | 0.0% | 7.0% | 50.0% |
| pm1 | 6 | 45 | 6 | 35 | 16 | 6 | 30 | 13 | 0.0% | 33.3% | 0.0% | 14.3% | 18.7% |
| cm150a | 9 | 46 | 9 | 46 | 20 | 6 | 37 | 10 | 33.3% | 19.5% | 33.3% | 19.5% | 50.0% |
| mux | 9 | 46 | 9 | 46 | 12 | 5 | 35 | 4 | 44.4% | 23.9% | 44.4% | 23.9% | 66.6% |
| i1 | 6 | 41 | 6 | 36 | 12 | 6 | 32 | 4 | 0.0% | 21.9% | 0.0% | 11.1% | 66.6% |
| decod | 3 | 28 | 3 | 28 | 6 | 3 | 28 | 4 | 0.0% | 0.0% | 0.0% | 0.0% | 33.3% |
| pcle | 8 | 67 | 8 | 62 | 17 | 7 | 48 | 18 | 12.5% | 28.3% | 12.5% | 22.6% | -5.0% |
| tcon | 2 | 24 | 2 | 24 | 8 | 2 | 24 | 1 | 0.0% | 0.0% | 0.0% | 0.0% | 87.5% |
| cc | 5 | 44 | 5 | 43 | 8 | 5 | 36 | 8 | 0.0% | 18.1% | 0.0% | 16.3% | 0.0% |
| Average reductions | | | | | | | | | 14.9% | 31.6% | 10.5% | 16.8% | 33.5% |

**Appendix**

In this section, the obtained circuits of applying the proposed method to 20 MCNC benchmarks are illustrated in Table A1.

Table A1: The obtained circuits of applying the proposed method to 20 MCNC benchmarks.

| Benchmarks | circuit |
|---|---|
| b1 | f=M(M(a,b',1)),M(a',c,0),M(b,c',0)), e=M(M(a,b',0),M(a,b',1)',1), d=c , g=c' |
| cm82a | f=M(M(a,b,c)',M(a,b',c),b), g=M(M(M(a,b,c),d,e)',M(M(a,b,c),d',e),d), h=M(M(a,b,c),d,e) |
| majority | f=M(d,M(M(a,b,M(c,e,0)),M(e,c,1),0),1) |
| 9symml | <52>=M(M(M([7815],M([7819]',M(M([7816]',[7817],<0>),M([7818],[7820]',<0>),<1>),M([7816],[7818]',<0>)),M([7817]',M([7816]',[7820],<0>),M([7819],[7820]',<0>))),M([7815]',M(M([7816],M([7817],[7819]',<0>),M([7818]',[7819],<0>)),M([7817],M([7816]',[7820],<0>),M([7818],[7820]',<0>)),<1>),M(M([7817]',[7818],<0>),[7819],<0>)),<1>),M(M(M([7817]',[7820],<0>),M(M([7816],[7818],<0>),[7819]',<0>),M([7818]',[7819],<0>)),<br>M(M(M([7816]',[7817],<0>),M([7819],[7820]',<0>),<0>),[7818]',<0>),<1>),<1>),<br>[7815]=M(1,2,3),<br>[7816]=M(M(1,M(2,3,<0>),M(2',3',<0>)),M(1',M(M(2,3,<0>),M(2',3',<0>),1)',0),1),<br>[7817]=M(4,5,6),<br>[7818]=M(M(4,M(5,6,<0>),M(5',6',<0>)),M(4',M(M(5,6,<0>),M(5',6',<0>),1)',0),1),<br>[7819]=M(7,8,9),<br>[7820]=M(M(7,M(8,9,<0>),M(8',9',<0>)),M(7',M(M(8,9,<0>),M(8',9',<0>),1)',0),1) |
| x2 | <k>=M(i',1,M(h',j,1)), <L>=M((M(h,j,0),M(h,j,1)',i')', <m>=M(M(h',i',0),j',0),<br><n>=M(M(i',M(h,j,1),1),M(M(a',b',0),c',0)',1), <o>=M(M(M(i,j,0),h',1),g',1),<br><p>=M(M(M(M(M(a',b',0),c,0),M(M(i,j,0),h,0),M(h',i',0)),M(j',0,M(i',1,M(M(d,e',0),h,0))),1),M(g',1,M(M(f,h',0),M(i,j,0),0)),1),<br><q>=M(M(M(M(h,j,0),M(M(M(a',b',0),c',0),i',1),0),M(j',M(h',i',0),M(M(d,e,0),M(h,i,0),0)),1),M(g',1,M(M(f,h',0),M(i,j,0),0)),1) |
| cm152a | M(M(i',M(k',M(a,j',0),M(c,j,0)),M(k,M(e,j',0),M(g,j,0))),M(i,M(k',M(b,j',0),M(d,j,0)),M(k,M(f,j',0),M(h,j,0))),1) |
| cm85a | L=M(M(M(M(M(h',i,M(j',k,0)),f',g),d',e),b,0),a,1),<br>n=M(M(b,0,M(M(M(h,i',M(j,k',0)),f,g'),d,e')),c,1),<br>m=M(b,0,M(M(M(M(M(h',i,M(j',k,0)),f',g),d',e),M(M(M(h,i',M(j,k',0)),f,g'),d,e'),1)') |
| cm151a | <m>=M(L',M(j,M(k,M(g,i',0),M(h,i,0)),M(k',M(c,i',0),M(d,i,0))),M(j',M(k,M(e,i',0),M(f,i,0)),M(k',M(a,i',0),M(b,i,0)))), <n>=<m>' |
| cm162a | <o>=M([556],f',1), <p>=M(M([559],f',1),[593],1), <q>=M(M([559],f',1),[627],1),<br><r>=M(M([559],f',1),[663],1), <s>=M(M(e,j,0),n,0), [45]=M(M(c,e,0),i',0),<br>[502]=M([45]',k,1), [554]=M(<s>,M(c,e,0)',1), [555]=M([45],M([554],i,0),1),<br>[556]=M([555],d,0),M(a,d,1)',1), [559]=M(<s>,M(c,d,0),0),<br>[592]=M(M([45]',k,1)',M([45]',k,0),1), [593]=M(M([592],d,0),M(b,d,1)',1),<br>[627]=M(M(M([502],L,0),M([502],L,1)',d),M(d,g,1)',1),<br>[663]=M(M(M(m,M([502],L,1),0),M(M([502],L,1),m,1)',d),M(d,h,1)',1) |
| cu | <p>=M(M(M(c,e',0),M(c',f',0),1),M(M(e,f,0),d,1),1),<br><q>=M([1402],M([1404],M(c,f',0),0),1), <r>=M(M(<w>,a',0),b',0),<br><s>=M(M(<w>,a,0),b',0), <t>=M(M(<w>,a',0),b,0), <u>=M(M(<w>,a,0),b,0),<br><v>=M([1404],[1398],M(M([1403],i',0),M(a,M(b,m',0),M(b,k,1)'),M(a',M(b,L',0),M(b',j,0)))), <w>=M([1402],o',0), <x>=M([1404],[1398],[1403]), <y>=M(g,o,0),<br><z>=M(M(d',g,0),M(c',f',1),0), [1398]=M(M(c,f',0),o',0),<br>[1402]=M(M(c',f,0),M(d',e',0),0), [1403]=M(M(c',f,0),M(n',o,0),0), [1404]=M(d',e,0) |
| cm163a | <q>=M(f,M(a,e',0),M(e,M(j,L0',0),M(j',L0,0)))',<br><r>=M(f,M(b,e',0),M(e,M(L,m0,0),M(L,m0,1)'))',<br><s>=M(f,M(g,e',0),M(e,M(m,q0,0),M(m,q0,1)'))',<br><t>=M(f,M(e',h,0),M(e,M(n,r0,0),M(n,r0,1)'))', <u>=M(d,p0',0),<br>p0=M(M(i,k,0),M(o,p,0),0)', r0=M(M(j,L0,1),M(L,m,1),1)', q0=M(M(j,L0,1),L,1)',<br>m0=M(j,L0,1)', L0=M(c,d,0)' |
| cmb | <q>=M(M(M(M(a,b,0),M(c,d,0),0),M(M(e,f,0),M(g,h,0),0),0),M(M(i,j,0),M(k,L,0),0),0),<br><r>=M(M(M(M(e,f,1),M(g,h,1),1),M(M(i,j,1),M(k,L,1),1),1),M(M(m,n,1),M(o,p,1),1),1),<br><s>=M(M(M(M(e,f,0),M(g,h,0),0),M(M(i,j,0),M(k,L,1),0),0),M(M(m,n,0),M(o,p,0),0),0)'<br>,<br><t>=M(M(M(M(e,f,1),M(g,h,1),1),M(M(i,j,1),M(k,L,1),1),1),M(M(m,n,1),M(o,p,1),1),1)' |
| pm1 | <r>=M(b,M(m,n,1),1), <s>=M(m',n,0)',<br><t>=M(M(M(k,n,0),m,0),M(M(g,h,0),M(i,j,0),0),0)',<br><u>=M(M(M(k,n,0),m,0)',M(M(g,h,0),M(i,j,0),0),1), <v>=p', <w>=o',<br><x>=M(M(b,k,0),M([2897],M(m',n,0),1),0),<br><z>=M(M(a',M(M(M(c,d,0),e,0),M(k,n,0),0),1),M(L,m',1),1),<br><a0>=M(M(a,L',0),M(m,n,0),M(m,n,1)') <b0>=M(M([2897],[2901],0),M(k,n,0),0)<br><c0>=M(M(<b0>,b,0),M([2901],M(b,n,1)',0),1), <d0>=M([2901],M(k',n,0),0),<br>[2897]=M(M(c,d,0),e,0)', [2898]=M(M(k,n,0),m,0)', [2901]=M(M(a,L',0),m,0) |
| cm150a | <v>=M(u',f1,f2)',<br>f1=M(t',M(s',M(r',M(a,q',0),M(b,q,0)),M(M(M(c',q',0),M(d',q,0),1)',r,0)),M(s,M(r',M(e,q',0),M(f,q,0)),M(r,0,M(M(g',p',0),M(h',q,0),1)'))),<br>f2=M(t,0,M(M(M(r',M(i,q',0),M(j,q,0)),M(r,0,M(M(k',q',0),M(L',q,0),1)'),1),M(M(r',M(m,q',0),M(n,q,0)),M(r,0,M(M(o',q',0),M(p',q,0),1)'),s),M(s',1,M(M(r',M(m,q',0),M(n,q,0)),M(r,0,M(M(o',q',0),M(p',q,0),1)'),s)))) |
| mux | M(u,M(M(q,M(M(s,t,0),M(a,r,0),M(e,r’,0)),M(M(s,t’,0),M(b,r,0),M(f,r’,0))),M(q,M(M(s’,t,0),M(c,r,0),M(g,r’,0)),M(M(s’,t’,0),M(d,r,0),M(h,r’,0))),1),M(M(q’,M(M(s,t,0),M(i,r,0),M(m,r’,0)),M(M(s,t’,0),M(j,r,0),M(n,r’,0))),M(q’,M(M(s’,t,0),M(k,r,0),M(o,r’,0)),M(M(s’,t’,0),M(l,r,0),M(p,r’,0))),1)) |
| i1 | <V27_1>=M(M((in_V27_0)',(in_V29_0),0),M([33],M(V8_0,V9_0,0),M((V8_0)',(V9_0)',0))),1),<br><V27_2>=M(M(M((in_V27_0),(in_V29_0),0),M(M(M(V7_1,V7_2,1),M(V7_3,V7_4,1),1),M(M(V7_5,V7_6,1),V7_7,1),1),0),M(M(V8_0,(V9_0)',0),[33],0),1),<br><V27_4>=M(in_V27_3,V22_2,1), <V28_0>=M(V10_0,M(V8_0',[33],0),1),<br><V30_0>=M(V18_0,V22_5,0), <V32_0>=M(V11_0,V22_5,0),<br><V33_0>=M(V14_0,M(V22_3,(V22_5)',0),0),<br><V34_0>=M(V17_0,M(V22_3,(V22_5)',0),0),<br><V35_0>=M(V14_0,M(V22_4,(V22_5)',0),0),<br><V36_0>=M(V17_0,M(V22_4,(V22_5)',0),0), <V37_0>=M(V16_0,(V22_5)',0),<br><V38_0>=M(M(V12_0,V13_0,1),M(V14_0,V15_0,1),1), <V27_0>=in_V27_0,<br><V27_3>=in_V27_3, <V29_0>=in_V29_0, <V31_0>=V11_0,<br>[33]=M(in_V29_0,M(M(M(V7_1,V7_2,1),M(V7_3,V7_4,1),1),M(M(V7_5,V7_6,1),V7_7,1),1)',0) |
| decod | f=M(M(a,d,0),M(e,M(b,c,0),0),0), g=M(M(a,d',0),M(e,M(b,c,0),0),0),<br>h=M(M(a,d,0),M(e,M(b,c',0),0),0), i=M(M(a,d',0),M(e,M(b,c',0),0),0),<br>j=M(M(a,d,0),M(e,M(b',c,0),0),0), k=M(M(a,d',0),M(e,M(b',c,0),0),0),<br>L=M(M(a,d,0),M(e,M(b',c',0),0),0), M=M(M(a,d',0),M(e,M(b',c',0),0),0),<br>N=M(M(a',d,0),M(e,M(b,c,0),0),0), O=M(M(a',d',0),M(e,M(b,c,0),0),0),<br>P=M(M(a',d,0),M(e,M(b,c',0),0),0), Q=M(M(a',d',0),M(e,M(b,c',0),0),0),<br>R=M(M(a',d,0),M(e,M(b',c,0),0),0), S=M(M(a',d',0),M(e,M(b',c,0),0),0),<br>T=M(M(a',d,0),M(e,M(b',c',0),0),0), U=M(M(a',d',0),M(e,M(b',c',0),0),0) |
| pcle | <t>=M(M(S,V0,0),U0,0), <u>=M(M(a,i,0),M(L',U0,0),1),<br><v>=M(M(b,i,0),M(M(L,m',0),M(L',m,0),U0),1),<br><w>=M(M(c,i,0),M(M(a1,n',0),M(a1',n,0),U0),1),<br><x>=M(M(d,i,0),M(M(o,Z0',0),M(o',Z0,0),U0),1),<br><y>=M(M(e,i,0),M(U0,M(M(p,M(o,Z0,0),0),M(p',M(o,Z0,0)',0),1)',0),1),<br><z>=M(M(f,i,0),M(M(q,x0',0),M(q',x0,0),U0),1),<br><a0>=M(M(g,i,0),M(U0,M(M(r,M(q,x0,0),0),M(r',M(q,x0,0)'),1)',0),1),<br><b0>=M(M(h,i,0),M(U0,M(S,v0',0),M(s',v0,0)),1), U0=M(M(i',j,0),k',0),<br>V0=M(M(r,q,0),x0,0), a1=M(L,m,0), Z0=M(a1,n,0), x0=M(M(o,p,0),Z0,0), |
| tcon | <a0>=M(a,M(i,k,1),M(i',k,0)), <b0>=M(b,M(i,L,1),M(i',L,0)),<br><c0>=M(c,M(i,m,1),M(i',m,0)), <d0>=M(d,M(i,n,1),M(i',n,0)),<br><e0>=M(e,M(i,o,1),M(i',o,0)), <f0>=M(f,M(i,p,1),M(i',p,0)),<br><g0>=M(g,M(i,q,1),M(i',q,0)), <h0>=M(h,M(i,r,1),M(i',r,0)), <s>=k, <t>=L, <u>=m,<br><v>=n, <w>=o, <x>=p, <y>=q, <z>=r |
| cc | <w>=M(L,v,0), <x>=M(M(M(i,k,0),q,0),p',0),<br><y>=M(M(L',m,0),M(M(M(i,k,0),q',0),p,1),0), <z>=M(<x>,m,0), <a0>=t', <f0>=M(i,j,0),<br><g0>=M(i,j,0)', <i0>=M(m,M([84],o,0),M(a,M(M(i,k,0),q',0),0)),<br><j0>=M(m,M(M(p,M(b,M(i,k,0),0)',0),M(M(i,k,0),q,0),1),0),<br><k0>=M(m,M(c,M(M(i,k,0),q',0),0),M(q,M(i,k,0)',0)),<br><L0>=M(m,M(M([84],p',0),r,0),M(d,0,M(M(i,k,0),q',0))), |
| | <m0>=M(m,M([84],s,0),M(e,0,M(M(i,k,0),q',0))),<br><n0>=M(m,M([84],t,0),M(f,0,M(M(i,k,0),q',0))),<br><o0>=M(m,M([84],u,0),M(g,0,M(M(i,k,0),q',0))),<br><p0>=M(m,M([84],v,0),M(h,0,M(M(i,k,0),q',0))), <b0>=u, <c0>=q, <d0>=s, <e0>=r,<br><h0>=p, [84]=M(M(i,k,0),q',0)' |